# Optical vortex generation from molecular chromophore arrays

Mathew D. Williams, Matt M. Coles, Kamel Saadi, David S. Bradshaw and David L. Andrews*

*School of Chemistry, University of East Anglia, Norwich NR4 7TJ, United Kingdom*
*\* To whom correspondence should be addressed.  E-mail: d.l.andrews@uea.ac.uk*

**Abstract:**  The generation of light endowed with orbital angular momentum, frequently termed optical vortex light, is commonly achieved by passing a conventional beam through suitably constructed optical elements.  This Letter shows that the necessary phase structure for vortex propagation can be directly produced through the creation of twisted light from the vacuum.  The mechanism is based on optical emission from a family of chromophore nano-arrays that satisfy specific geometric and symmetry constraints.  Each such array can support pairs of electronically delocalized doubly degenerate excitons whose azimuthal phase progression is responsible for the helical wave-front of the emitted radiation.  The exciton symmetry dictates the maximum magnitude of topological charge; detailed analysis secures the conditions necessary to deliver optical vortices of arbitrary topological charge.

**Keywords:** Optical angular momentum, structured light, optical vortex, quantum optics, quantum electrodynamics, nano-array, chirality

**PACS:**  03.70.+k; 42.50.Ct; 42.50.Tx; 78.67.Pt; 81.07.Nb

Light with a twisted wave-front is distinguished by the topology and azimuthal phase variation in its optical fields, and an associated orbital angular momentum [1].  From the earliest experiments [2, 3] a wide range of innovative applications has emerged – including, but not limited to: micro-particle rotary manipulation [4], new strategies for the entanglement of optical states [5, 6], the transmission of quantum information [7], and the use of associated optical elements for image enhancement [8, 9].  While the fundamental photonic properties of such 'optical vortex' beams are now well established, this is also a field that is generating new discoveries at the fundamental level, some inviting reconsideration of long-held tenets such as how much information can be conveyed or measured in an individual photon [10].  One of the key principles is that a beam with a topological charge $l$ (an integer signifying the number of wave-front twists per unit wavelength) propagates in the form of photons that convey an orbital angular momentum $l\hbar$, in addition to any spin angular momentum relating to polarization state.

Historically, the generation of vortex radiation is accomplished by active conversion of a conventional optical beam such as with spatial light modulators [11, 12], a q-plate [13] or a spiral phase plate [14].  Such methods typically produce an output containing more than one vortex mode, with positive and negative topological charges separable through angular filtering [15, 16].  It has not generally been considered possible to directly generate optical vortex radiation with a specific, pre-determined topological charge, by engaging the electronic processes of source emission.

To secure such a form of output requires a form of correlation between the spatial and electronic properties of the emitters that is impossible to achieve with the atomic components of any conventional source.  However, we now demonstrate that the specific conditions for the direct production of vortex emission can be satisfied in molecular arrays of specified geometry and symmetry.  Moreover, the tailored design of these arrays provides for generating vortices of arbitrary topological charge.

To begin, consider an organized molecular array comprising $n$ elements of identical chemical composition, each with its own electronic integrity.  The components are rigidly held by a host material – close enough together to experience Coulombic interactions, but not enough for significant overlap between their electronic wavefunctions.  The molecular composition of each component is such that it can act as a chromophore/fluorophore, displaying characteristic absorption and emission processes in the visible wavelength region.  A wide range of such materials has been synthesized; in many cases the chromophores are sub-units of a larger molecular structure playing the roles of both a support and a dielectric insulator between the individual chromophores.  While the electronic ground state wavefunction of such an array essentially comprises a product of the ground state wavefunctions of the individual components, the electronic excited states are delocalized around the ring and experience a field splitting associated with the lowering of symmetry due to nearest-neighbor interactions.

To achieve the objective of vortex emission it is expedient to focus on cylindrically symmetric, ring-like arrays of $n \geq 3$ chromophores with non-parallel transition dipoles, centered upon an axis of rotational symmetry of order $n$. This is a configuration that confers precisely equivalent electronic coupling between each neighboring pair, assuming negligible interaction with the surrounding host material. The Schoenflies point group symmetry designation of such a system is $C_n$ or $C_{nh}$ [17], depending on the absence or presence, respectively, of reflection symmetry in the plane that includes the centers of each member of the array. For simplicity we adopt the former, more general, case (signifying 3D, rather than planar, chirality); this removes a restriction for the chromophore emission dipoles to lie in the plane.

To determine the excited electronic states that can be supported by such an array it is expedient to block diagonalize the Hamiltonian matrix for the system. This procedure elicits wavefunctions for excited states in which the electronic energy is substantially delocalized around the ring; in standard terminology these are referred to as excitons. Generally, excitons emerge in various symmetry forms, each distinct symmetry being associated with a particular energy. Our focus will be on the wavefunctions for specific forms of exciton, to be indexed with a numerical integer $p$. These states, corresponding to the lowest electronically excited levels of the array, comprise superpositions of $n$ basis states – in each of which one chromophore is excited and the others are in the ground state. In Dirac notation, the exciton state vectors are thus expressible as follows:

$$|\psi_p\rangle = \frac{1}{\sqrt{n}} \sum_{r=1}^{n} \varepsilon_n^{(r-1)p} |\xi^{r;u}\rangle \prod_{s \neq r} |\xi^{s;0}\rangle. \quad (1)$$

Here $|\xi^{r;u}\rangle$ is the state vector for chromophore $r$ in its excited state $u$, $|\xi^{r;0}\rangle$ represents the corresponding ground state, and $\varepsilon_n = \exp(2\pi i/n)$. The number of exciton states is equal to the order of the point group, again simply $n$ in the case of $C_n$ symmetry.

The exciton states fall into two distinct symmetry categories: there are $\lfloor (n-1)/2 \rfloor$ (signifying the largest integer not greater than $(n-1)/2$) doubly degenerate pairs with Schoenflies irreducible representations labeled $E_q$, with $1 \leq q \leq \lfloor (n-1)/2 \rfloor$; also there is one totally symmetric, non-degenerate state with symmetry label $A$. Furthermore, in the case of even $n$, there is an additional non-degenerate state of $B$ symmetry. The $A$ states arise from $p = n$, and in consequence do not exhibit a phase progression around the ring, since in Eq. (1) the factor $\varepsilon_n^{(r-1)p}$ then equals unity. Equally it is evident that the phase progression of $B$ states, where $p = n/2$, lacks specific circularity. Attention is therefore directed to the $E_q$ states, for each of which the doubly degenerate components display circular phase progressions of equal topological charge $q$, with opposite signs. For example, in the $E_2$ exciton state of a five-member ring comprising chromophores $r = \{1, 2, \ldots, 5\}$, each electronically excited chromophore carries a phase factor of $\exp(\pm 4\pi i r/5)$, consistent with a topological charge of $\pm 2$; this is illustrated by Fig. 1.

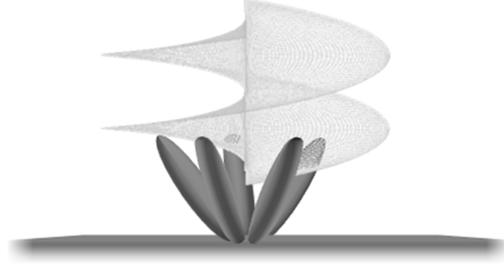

Fig. 1. Schematic depiction of five oriented chromophores and the ensuing optical vortex emission of topological charge $l = 2$, from the decay of an associated delocalized exciton: the helix depicts a surface of constant optical phase.

Next, we consider the radiative transitions that occur as the exciton states decay. Notably, the different symmetry designations of excitons with symmetries $E_q$, $A$ (and $B$, if present) are reflected in their association with different energies. Each of the double degenerate excitons labeled $q$ will therefore produce the emission of a photon with a characteristic wavelength, corresponding to the initial energy level. To identify the energy levels, the Hamiltonian operator can be cast in matrix form as $\hat{H}_{rs} = \delta_{rs} E_u + (1-\delta_{rs})\hat{V}_{rs}$, in which the second term denotes electrodynamic coupling between dipoles. The energy eigenvalues emerge in the form $E_u + 2V_u \cos(2\pi q/n)$, where

$$V_u \equiv V_{r,(r+1)\bmod n}\left(k_u, \mathbf{R}_{r,(r+1)\bmod n}\right) \equiv V_{12}(k_u, \mathbf{R}_{12})$$
$$= \frac{e^{ik_u R}}{4\pi\varepsilon_0 R^3}\Big[\{1 - ik_u R - (k_u R)^2\}(\boldsymbol{\mu}_1^{0u} \cdot \boldsymbol{\mu}_2^{0u})$$
$$- \{3 - 3ik_u R - (k_u R)^2\}\{(\boldsymbol{\mu}_1^{0u} \cdot \hat{\mathbf{R}}_{12})(\boldsymbol{\mu}_2^{0u} \cdot \hat{\mathbf{R}}_{12})\}\Big]. \quad (2)$$

This result demonstrates that each nearest-neighbor interaction produces a coupling of identical magnitude to that which exists between chromophores 1 and 2, as befits the ring symmetry. In Eq. (2), the nearest neighbor distance $R$ is defined by $\mathbf{R}_1 - \mathbf{R}_2 \equiv \mathbf{R}_{12} = R\hat{\mathbf{R}}_{12}$, the carat here denoting a unit vector; the individual electric dipole transition moments are defined by $\boldsymbol{\mu}_r^{0u} \equiv \langle \xi^{r;0} | \boldsymbol{\mu}^{(r)} | \xi^{r;u} \rangle$ and $k_u = E_u/\hbar c$.

In consequence of the $q$-dependence of the energy levels, it is possible to selectively excite an individual $E_q$ state with a topological charge of specific magnitude. To identify the optical field produced by the ensuing exciton decay requires summing the retarded electric field components associated with each emitter [18], accounting for their relative phases as given by Eq. (1). The result is as follows:

$$\mathcal{E}_p(\mathbf{R}_D) = \sum_r^n \frac{e^{ikR_{Dr}} \varepsilon_n^{(r-1)p}}{4\pi\varepsilon_0 R_{Dr}^3} \left\{ \left[ \left(\hat{\mathbf{R}}_{Dr} \times \boldsymbol{\mu}_r^{0u}\right) \times \hat{\mathbf{R}}_{Dr} \right] k^2 R_{Dr}^2 \right. \\ \left. + \left[ 3\hat{\mathbf{R}}_{Dr} \left(\hat{\mathbf{R}}_{Dr} \cdot \boldsymbol{\mu}_r^{0u}\right) - \boldsymbol{\mu}_r^{0u} \right] (1 - ikR_{Dr}) \right\}, \quad (3)$$

expressing the field detected at a position $\mathbf{R}_D$. From this result we can readily identify the phase distribution of the associated optical field, $\theta_i(\mathbf{R}) = \arg\{E_{p;i}(\mathbf{R}_D)\}$, where the individual Cartesian components, indexed by $i$, in the case of a three-particle ring, are explicitly exhibited in the Supplementary Material. Fig. 2 shows color-coded representations [19] of the phase of the field emitted in the $z$-direction, normal to the plane containing the ring at $z = 0$, and relating to the right-handed component of the vortex pair. The three Cartesian electric field components exhibit identical phase progression in the far field. For vortex propagation in 3D space, there has to be a phase singularity on a 1D contour, which is achievable with three or more emitters. The illustrations in Figs 2 (*a*)-(*d*) exhibit nano-arrays with differing numbers of chromophores, from which twisted light is emitted with a variety of topological charges.

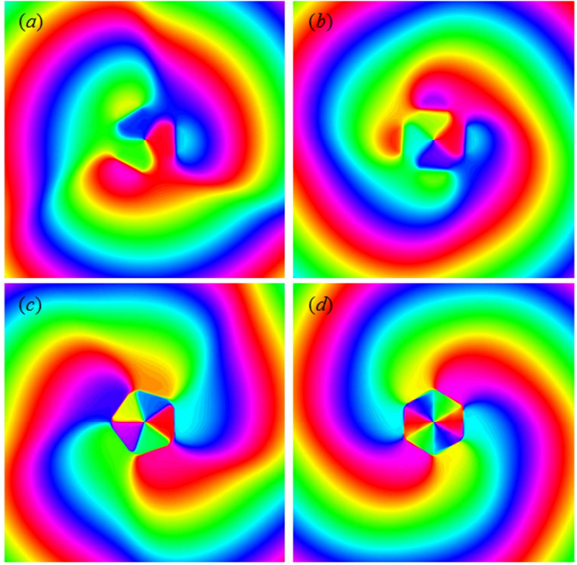

Fig. 2. Sample phase maps ($i = z$) for the optical vortex emission from *n*-chromophore rings in the direction normal to the plane, distributed about the center of each array: (*a*) $n = 3$ ($l = +1$); (*b*) $n = 4$ ($l = -1$); (*c*) $n = 5$ ($l = -2$); (*d*) $n = 6$ ($l = +2$). The angle subtended by each emission dipole moment onto the plane containing the array centers is $\pi/4$, and the angle between each planar projection and the radial position vector of the corresponding emitter is also $\pi/4$. In this case the coupling constant $V_u$ is positive.

First, we observe that emission in either direction perpendicular to the plane of the nano-array will have the opposite sense of wave-front helicity to the vortex emitted by decay of the same exciton in the opposite direction. In practice, attachment of the chromophore ring to a surface will favor outward emission. As noted earlier, the case of a threefold array, exhibited in Fig. 2 (*a*), represents the lowest order multiplicity from which an optical vortex can be generated. This is consistent with the premise that it requires the interference of three or more plane waves to create an axis of phase singularity [20]. Plots (*a*) and (*b*) both exhibit the optical phase winding about the *z*-axis, with a complete $2\pi$ phase shift resulting from complete rotation. The counterclockwise rotation in (*a*) is consistent with a positive topological charge, here $l = +1$, while the phase progression in (*b*) has a clockwise sense, $l = -1$. Figs 2 (*c*) and (*d*) show the $4\pi$ shift shifts associated with $l = \pm 2$, the highest value achievable with five emitters ($C_5$ symmetry). Emission with $l = \pm 1$ (not illustrated) can also be achieved with five emitters; indeed it is possible to separately excite either case, exploiting the fact that the corresponding array excited states will differ in energy.

Reflecting on the possibility of securing vortex emission of a specific handedness, there is an obvious requirement to remove the degeneracy between the relevant pair of *E* states, so that one component can be more selectively excited by appropriate choice of excitation wavelength. The result can be achieved by undermining the (*x*, *y*) symmetry responsible for the twofold state degeneracy, as for example might be achieved by anchoring a surface array to an anisotropic substrate.

Another interesting observation is that the multi-chromophore mechanism, for the generation of optical vortices with arbitrary topological charge, is consistent with the array as a whole supporting an electronic transition of *multipole* character, notwithstanding the electric *dipole* character of the electronic transition on each emitter. It is readily verified that the doublet $E_q$ states in systems of $C_n$ symmetry support electric multipole transition moments of order $w$ constrained by $w \geq q$. The result throws light on the long-sought linkage between the order of emission multipolarity and photon angular momentum [21-23].

The challenge is to fabricate the kinds of material that can support vortex emission. Certainly, a number of nanostructured materials already synthesized and characterized have very similar chemical, physical and symmetry properties to those that are now required. Perfluorodiphenylborane provides one of the simplest examples of a chiral chromophore array satisfying the condition of threefold symmetry, its own chiral character resulting from steric hindrance. Notably, the key symmetry requirements are only fulfilled with molecular chromophores, not atomic arrays – since the latter would introduce planes of mirror symmetry undermining the necessary chirality. As we have indicated, the simplest system could be a nanofabricated surface suitably tailored with preformed, surface-deposited molecular arrays.

One might further suppose that metastructured materials based on such units could be formed into stacks within the cavities formed by a porous silicate, zeolite or similar lattice, or in an electrically poled liquid crystalline environment. Consideration would have to be given to achieving suitable phase-matching along the emission direction. Bulk materials of such a kind might then provide the basis, given a suitably constructed optical cavity and pump mechanism, to create a viable optical vortex laser source.

Before concluding, we have to draw attention to the remarkable fact that exactly the right kind of chromophore ring structures commonly occur in the natural photosynthetic apparatus of plants and photobacteria [24, 25]. In native form, however – at least in green plant systems – they do not have the capacity to form the necessary excitons of specific symmetry class, due to local breaking of the intrinsic array symmetry by the complex electronic environment of a less highly symmetric protein host.

In conclusion, the principles that we have established offer scope for the development of a vortex laser, enabling radiation with a helical wave-front to be tailor-made by creation from the vacuum, instead of through modifications to the emission from a regular laser source. In addition to its fundamental interest, such a system also has the capacity to entirely circumvent conventional methods normally requiring the passage of laser light through spiral phase plates or other optical elements having the same effect. Pursuing this concept, in following work we plan to tackle issues associated with creation of a vortex beam with polarization varying in the plane perpendicular to the direction of propagation. A preliminary analysis points to the emission of radiation with polarization varying over the beam profile, as produced in recent experiments using q-plates [13, 26]. We also plan to explore connections with established local measures of electromagnetic chirality [27, 28].

This research is supported by funding from the Leverhulme Trust and the Engineering and Physical Sciences Research Council.

# Optical vortex generation from molecular chromophore arrays

Mathew D. Williams, Matt M. Coles, Kamel Saadi, David S. Bradshaw and David L. Andrews

*School of Chemistry, University of East Anglia, Norwich Research Park, Norwich NR4 7TJ, United Kingdom*

**Supplementary material**

### S1. Calculational basis

Equation (3) is the starting point, expressing the fully retarded electric field generated from each emitting transition dipole $\mu_r^{0u}$ and detected at position $\mathbf{R}_D$;

$$\mathcal{E}_p(\mathbf{R}_D) = \sum_r^n \frac{e^{ikR_{Dr}} \varepsilon_n^{(r-1)p}}{4\pi\varepsilon_0 R_{Dr}^3} \left\{ \left[ \left( \hat{\mathbf{R}}_{Dr} \times \mu_r^{0u} \right) \times \hat{\mathbf{R}}_{Dr} \right] k^2 R_{Dr}^2 + \left[ 3\hat{\mathbf{R}}_{Dr} \left( \hat{\mathbf{R}}_{Dr} \cdot \mu_r^{0u} \right) - \mu_r^{0u} \right] (1 - ikR_{Dr}) \right\} , \quad (S.1)$$

From this, we develop a representative example used to derive the explicit form of the phase expression. In the following, a formula is found for a field generated by a set of *three* emitting sources having the C$_3$ propeller symmetry of the kind described in the main text. Since each electric dipole is orientated at ±120° with respect to its two adjacent dipoles in such a system, the electric fields from each emitter will add with the relevant phase factors {0, 2, 4}$\pi i$/3 so that, in a simplified form, the resultant is expressible as follows;

$$\begin{aligned}\mathcal{E}^{(ABC)} = \frac{e^{ikR}}{4\pi\varepsilon_0 R^3} &\left\{ \left[ (1-ikR)\{3\hat{\mu}_R^A \hat{\mathbf{R}}_A - \mu_A\} + \{\mu_A - \hat{\mu}_R^A \hat{\mathbf{R}}_A\} k^2 R^2 \right] \right. \\ &+ \left( -\frac{1}{2} + i\frac{\sqrt{3}}{2} \right) \left[ (1-ikR)\{3\hat{\mu}_R^B \hat{\mathbf{R}}_B - \mu_B\} + \{\mu_B - \hat{\mu}_R^B \hat{\mathbf{R}}_B\} k^2 R^2 \right] \\ &+ \left. \left( -\frac{1}{2} - i\frac{\sqrt{3}}{2} \right) \left[ (1-ikR)\{3\hat{\mu}_R^C \hat{\mathbf{R}}_C - \mu_C\} + \{\mu_C - \hat{\mu}_R^C \hat{\mathbf{R}}_C\} k^2 R^2 \right] \right\} , \quad (S.2)\end{aligned}$$

where $r \in \{A, B, C\}$ and $\hat{\mu}_R^r \equiv \hat{\mathbf{R}}_r \cdot \mu_r$. Each individual polarization component has an associated magnitude and phase, the separate cases being considered in Section S2.

### S2. Individual polarization components

The optical field analysis in the main text encapsulates the characteristic differences in the emission from excitons in arrays with various numbers of components, the main focus of the present Letter. The analysis also enables identification of counterpart expressions for $\theta$ considering the $x$, $y$ and $z$ components individually; for example:

$$\theta_i(\mathbf{R}) = \arg\{E_{p;i}(\mathbf{R}_D)\}, \qquad (\text{S.12})$$

where $i$ indexes the Cartesian components. Plotted in Fig. S1 are each of these three components for phase, again for the three-emitter case.

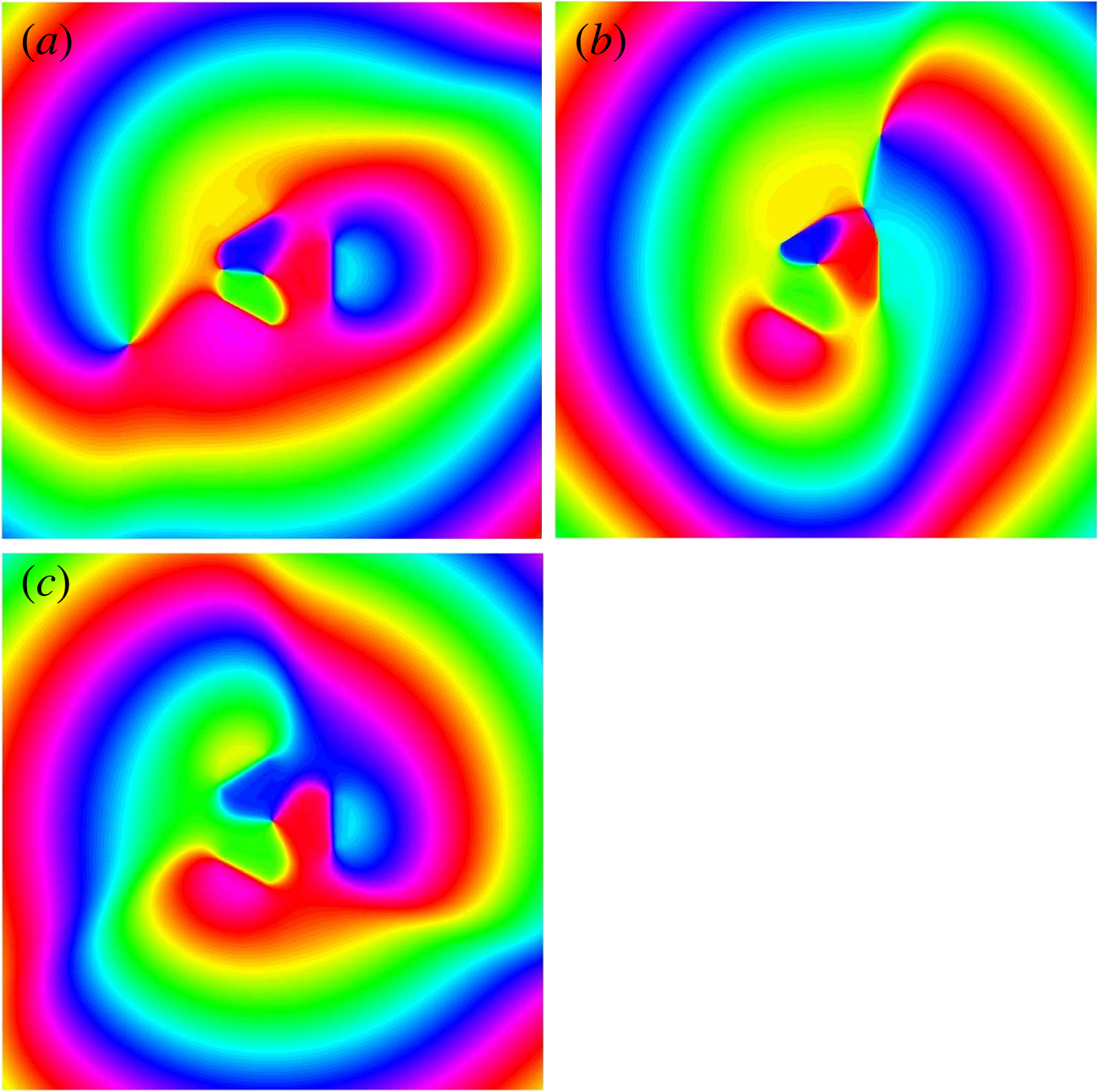

**Fig. S1** Phase maps of the individual polarization components for the case of three nanoemitters, disposed as in the caption for Fig. 2: (*a*) *x*-component; (*b*) *y*-component; (*c*) *z*-component.

## S3. Note on the cases of one or two emitters

The principal condition for generating excitons of the necessary symmetry properties cannot be satisfied with either $n = 1$ or 2 emitters; the reason is the upper limit of $\lfloor (n-1)/2 \rfloor$ on the number of doubly degenerate excitons of the necessary chiral form, as shown in the main paper. For comparison purposes, Fig. S2 shows the scalar phase maps for the single and double emitter cases; in the latter case there is a $\pi$-phase shift between the two emitters, generating a phase singularity on a 2D contour (*i.e.* the plane bisecting the emitter positions).

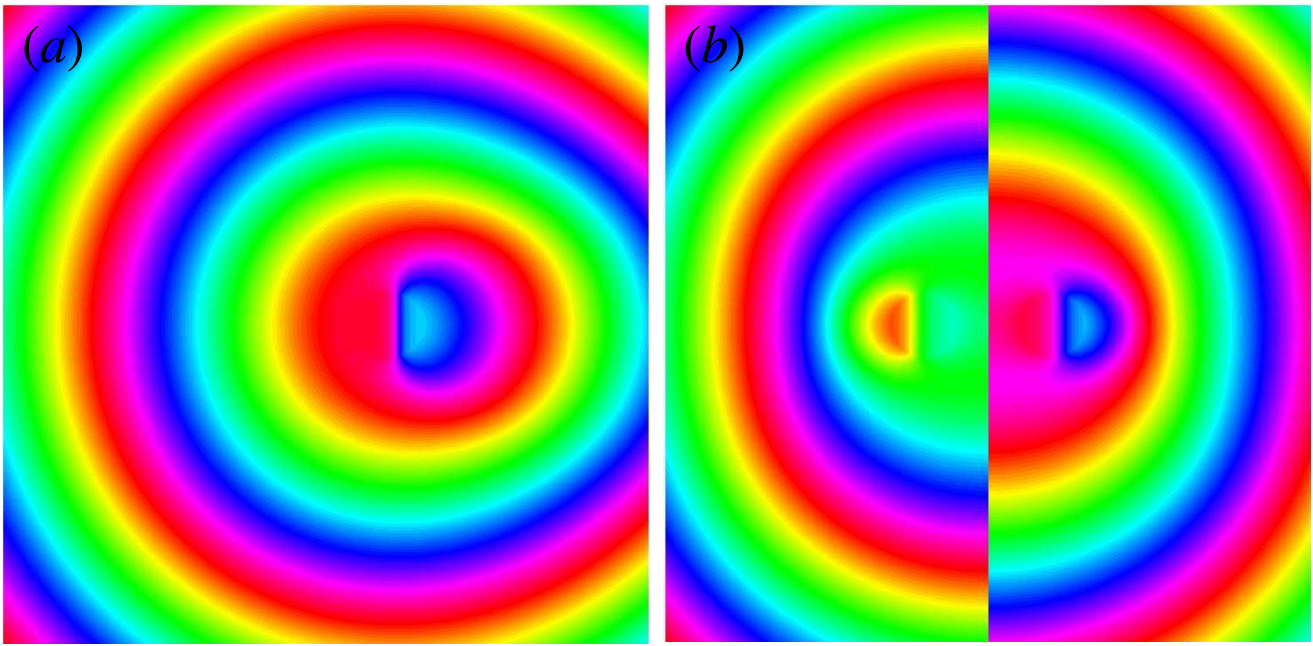

**Fig. S2**. Phase plots for (*a*) $n = 1$ and (*b*) $n = 2$ nanoemitters with $\pi$-phase shift (Note the lack of circular symmetry due to the dipole(s) orientation). There is no wavefront chirality in the emission.